ARTICLE

# Rapid local compression in active gels is caused by nonlinear network response

D. Mizuno[a], C. Tardin[b], and C. F. Schmidt[c]

Abstract

The actin cytoskeleton in living cells generates forces in conjunction with myosin motor proteins to directly and indirectly drive essential cellular processes. The semiflexible filaments of the cytoskeleton can respond nonlinearly to the collective action of motors. We here investigate mechanics and force generation in a model actin cytoskeleton, reconstituted *in vitro*, by observing the response and fluctuations of embedded micron-scale probe particles. Myosin mini-filaments can be modelled as force dipoles and give rise to deformations in the surrounding network of cross-linked actin. Anomalously correlated probe fluctuations indicate the presence of rapid local compression of the network that emerges in addition to the ordinary linear shear elastic (incompressible) response to force dipoles. The anomalous propagation of compression can be attributed to the nonlinear response of actin filaments to the microscopic forces, and is quantitatively consistent with motor-generated large-scale stiffening of the gels.

## 1. Introduction

High-magnification optical microscopy of living cells reveals vigorously fluctuating organelles, vesicles, filaments and other macromolecular structures [1]. Although thermal motions are not negligible for such small objects, most of the observed fluctuations can be attributed to non-equilibrium activity driven by metabolism. In cells, the most prominent forces are generated by motor proteins that interact with cytoskeletal filaments. The cytoskeleton is a viscoelastic network made of semiflexible polymers, that, on the one hand, confines embedded organelles and suppresses thermal diffusive migration, and, on the other hand, propagates motor-generated forces. These forces play central roles in intracellular material transport, cell migration and cell division. Cells transmit active tension through cytoskeletal networks when they adhere to extra-cellular matrices. Forces are also tied into physiological signaling processes [2].

Forces generated by motor proteins in isolation from other cytoplasmic constituents have been measured with single-molecule techniques [3]. However, the collective functions of motors and the cytoskeleton in the highly complex context of the cell remain obscure. Moreover, forces and mechanics are intricately related because of the highly nonlinear response characteristics of biological materials [4, 5]. Simultaneous observation of forces and mechanical properties in simple *in-vitro* actin-myosin networks that serve as models for the cytoskeleton help to elucidate the mechanical machinery of cells. Such measurements have become possible with progress in microrheology techniques [4, 6].

The term microrheology (MR) refers to a spectrum of techniques designed to probe the mechanical properties of small systems such as individual cells [7]. One-particle MR measures the local viscous and elastic properties of a material by observing the motion of individual embedded micrometer-sized probe particles [8-11]. Correlated movements of two widely-separated probe particles, being much less affected by the immediate environment and the surface chemistry of the probe particles, can be evaluated to determine larger-scale

[a.] Dept. Physics, Kyushu University, 819-0395 Fukuoka, Japan.
[b.] Institut de Pharmacologie et de Biologie Structurale, Université de Toulouse, CNRS, UPS, 31 077 Toulouse, France.
[c.] Dept. Physics and Soft Matter Center, Duke University, Durham, NC 27708, USA.

rheological properties (2-particle MR) [12-15]. Both 1-particle and 2-particle MR have been conducted as "passive microrheology" (PMR)[8-12], merely monitoring fluctuations of the probes and extracting the material response using the fluctuation-dissipation theorem (FDT) if fluctuations are thermal. Both techniques can also be used actively (AMR)[16-20] where one directly measures the response of micron-sized particles to an oscillating driving force, *e.g.* generated by optical trapping.

In thermodynamic equilibrium, AMR and PMR give results obeying the FDT [21], whereas the FDT is violated in samples far from equilibrium. Simultaneous measurements of both AMR and PMR, in that case, make it possible to quantify the degree of FDT violation under the assumption that thermal and non-thermal fluctuations are uncorrelated [4, 21]. Motions of probe particles (recorded by PMR) in an active soft material such as a cell are generally driven by both thermal and non-thermal forces. To study the active part of the dynamics, it is necessary to separate contributions from non-thermal sources out of the total (thermal + non-thermal) motions of probe particles. The averaged mean-squared thermal contributions can be estimated by measuring the mechanical response of the material with AMR and applying the FDT. The non-thermal contributions can then be obtained by subtracting the estimated thermal fluctuations from the total measured fluctuations. This approach allows on to analyze active force generation in non-equilibrium media [21, 22].

The mechanical properties of active materials are, in general, changed by applied stresses [4]. Therefore it is not possible to use an equilibrium control experiment, *e.g.* observing a de-energized system, to obtain the relevant mechanical properties of an active material. It is necessary to measure the mechanical response directly in the active material [21-23]. Once the spatial distribution of mechanical properties is known, one can start to investigate the propagation of active forces in the material [24].

Here, we employed 1-particle AMR/PMR and 2-particle PMR in *"active gels", i.e.* reconstituted actin networks driven by myosin motor proteins. We used 2-particle PMR to measure characteristic length scales of propagation of motor-generated stresses and strains. We evaluated statistical averages of probe correlations as a function of particle distance. Both, thermal fluctuations in crosslinked non-active actin gels (without myosin) and non-thermal fluctuations in active gels did not show a anomalous length-scale dependency. However, fluctuations of pairs of particles normal to the direction connecting them were significantly more strongly correlated than expected. This anomaly can be explained by the presence of local network compression or draining against the embedding solvent. We propose that the compression is caused by the nonlinear mechanical response of actin filaments to motor-generated forces [25]. The nonlinear response can also be observed through the global stiffening of the gel.

## 2. Methods

*Sample preparation*

Actin and myosin II were prepared from rabbit skeletal muscle according to published methods [26]. Actin was stored at -80°C in G-buffer (2 mM Tris-Cl, 0.2 mM $CaCl_2$, 0.5 mM DTT, 0.2 mM ATP, pH 7.5) and myosin at -80°C in high salt buffer (0.6 M KCl, 50 mM $KH_2PO_4$, pH 6.5). Actin was biotinylated following a standard procedure [27]. Unlabeled actin, and biotinylated actin were mixed with probe particles and neutravidin (Molecular Probes) as a crosslinker, and co-polymerized in F-buffer (2 mM HEPES, 2 mM $MgCl_2$, 50 mM KCl,1 mM EGTA, pH 7.5) which was pre-mixed with myosin and ATP (3.5 mM). The ratio of normal actin: biotinylated actin: neutravidin was 180:5:2. The total concentration of actin was 1 mg/ml and of myosin 170 nM. Unless stated otherwise, $2a$ = 1.16 µm silica particles were used as probes for microrheology by laser interferometry and 1.1 µm diameter polystyrene beads (Polysciences) were used for video microscopy. Crosslinking prevented macroscopic phase separation (superprecipitation) [28] while the samples still exhibited vigorous non-thermal fluctuations. For control experiments, crosslinked actin gels were prepared following the same protocol, but leaving out myosin.

*Microrheology*

Details of the optical-trap and laser-interferometry-based microrheology setup are given elsewhere [29, 4]. In AMR, the mechanical properties of the sample were determined from the response of embedded, micron-sized probe particles to imposed oscillatory forces. The particles were manipulated with a spatially oscillated optical trap (drive laser, $\lambda$ = 1064 nm, 4 W cw, Nd:YVO$_4$, Compass, Coherent Inc.), and the applied forces and displacements were detected by laser interferometry [29] with a second probe laser ($\lambda$ = 830 nm, 150 mW cw, IQ1C140/6017, Laser 2000) superimposed on the trap laser. For PMR, the oscillation of the drive laser is turned off in order to measure the spontaneous thermal and (if present) non-thermal fluctuations of a probe particle in the stationary optical trap. In order to be able to observe slow fluctuations of many embedded probe particles simultaneously, the sample was imaged by an ordinary bright-field microscope and videos were recorded at a 30 Hz frame rate. The probe particles' center-of-mass positions were obtained by particle tracking software (LabView) and thereafter analyzed in the same way as the laser interferometry data.

*Active and passive MR*

In AMR, the displacements $u^j(t) = u^j(\omega)\exp(i\omega t)$ of embedded, micron-sized probe particles to imposed forces

$f^i(t) = f^i(\omega)\exp(i\omega t)$ are measured with laser interferometry [29] in order to obtain the response functions $A^{ij}(\omega) = u^j(\omega)/f^i(\omega)$ [21]. Here the superscripts $i$ and $j$ denote the probe particle; $i = j = 1$ for 1-particle MR and $i = 1, j = 2$ for 2-particle MR. In the case of 2-particle MR, local coordinates are chosen following the symmetry of the system such that $x$ is parallel ($\parallel$) and $y$ is perpendicular ($\perp$) to the line connecting the particle centers of a given pair of particles. For an isotropic, homogeneous medium, two independent response parameters $A^{12}_\parallel(\omega) = u^2_x/f^1_x$ and $A^{12}_\perp(\omega) = u^2_y/f^1_y$ determine the full linear response. For PMR, the spontaneous bead displacements are measured, either in a stationary laser beam by laser interferometry or, without a laser, by video microscopy. Time-series data are Fourier transformed to calculate the power spectral densities, $C^{ij}_\parallel(\omega) \equiv \int_{-\infty}^{\infty} \langle u^i_x(t)u^j_x(0)\rangle e^{i\omega t}dt$, $C^{ij}_\perp(\omega) \equiv \int_{-\infty}^{\infty} \langle u^i_y(t)u^j_y(0)\rangle e^{i\omega t}dt$. When the sample is in thermodynamic equilibrium, this function is related to the imaginary part $A''(\omega)$ of the corresponding complex response function $A(\omega) = A'(\omega) + iA''(\omega)$ via FDT [30],

$$C(\omega) = \frac{2k_B T}{\omega} A''(\omega), \qquad (1)$$

where $k_B$ is the Boltzmann constant and $T$ is the temperature. Here and hereafter, subscripts ($\parallel, \perp$) and superscripts ($i, j, 12, etc.$) are omitted for simplicity when making general remarks.

## 3. Results

Fig. 1 a shows the 1-particle response functions of colloidal particles (diameter $2a = 1.16~\mu$m) in cross-linked actin/myosin active gels. Circles are the imaginary part of $A^{11}(\omega)$ measured with AMR, and the curve is derived from the probe particles' spontaneous fluctuations $\omega C^{11}(\omega)/2k_B T$ measured with PMR. A significant violation of the fluctuation dissipation theorem is observed at low frequencies while AMR and PMR agree well at high frequencies. FDT violation is observed only at low frequencies because the system is driven out of equilibrium by the slow, processive action of the active components (myosin mini-filaments). The thermal fluctuations can still be estimated from the material's response function $A''$ via FDT [30].

Our goal here is to investigate the statistical and dynamical characteristics of the non-thermal fluctuations in order to understand the subtle connections between mechanical properties and microscopic force generation in active gels. In order to achieve this, we first examine the mechanical properties of the active gel. With the assumption of an isotropic, homogeneous, and incompressible continuum, the response functions measured with MR are related to the shear modulus $G(\omega)$ as [13, 15]

$$A^{11}(\omega) = 1/6\pi G(\omega)a, \quad A^{12}_\parallel(\omega) = 2A^{12}_\perp(\omega) = 1/4\pi G(\omega)R. \quad (2)$$

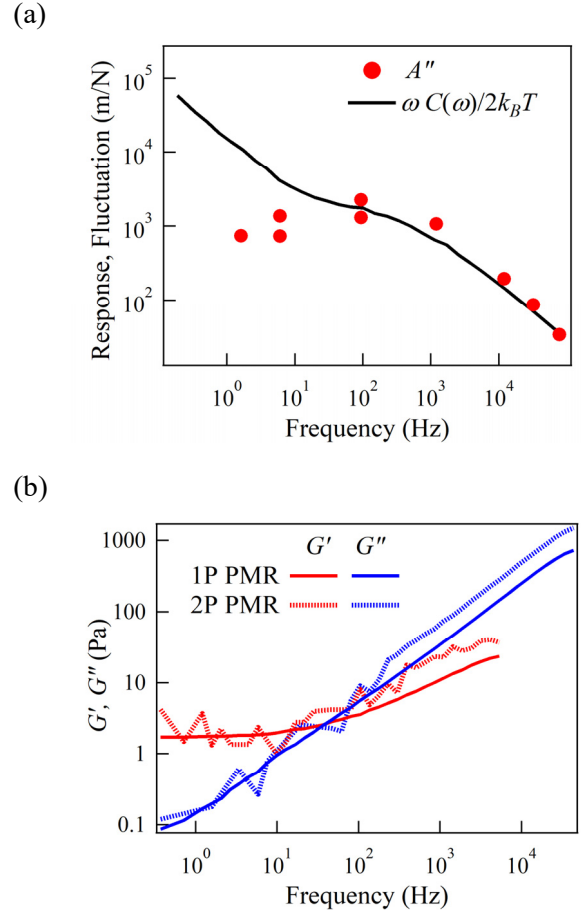

Fig. 1: (a) Active and passive 1-particle microrheology for a cross-linked active gel (probe particle : $2a = 1.16~\mu$m silica particles). Circles show $A''$ measured with AMR and the drawn curve shows the corresponding apparent response $\omega C(\omega)/2k_B T$ measured with PMR. The differences between the two curves at low frequencies indicate violation of the FDT. (b) Complex shear modulus measured with 1- and 2-particle PMR in cross-linked actin network without myosin (measured with 1.16 $\mu$m beads). Dashed and solid curves show 2-particle MR and 1-particle MR data, respectively.

These assumptions, however, cannot *a priori* be made for a network of actin filaments. They could be questioned because of the long persistence length of actin filaments of ~10 $\mu$m [31, 32]. One-particle MR is in such a case expected to provide local mechanical properties, which do not necessarily have to agree with larger-scale material response. For entangled (*i.e.* non-crosslinked) actin networks, it has indeed been observed that the mechanical response is length-scale dependent; $G(\omega)$ estimated from 1-particle MR using Eq. (2) is dependent on the probe particle size, and is inconsistent with the shear modulus obtained from 2-particle MR, or macrorheology [12, 33]. In Fig. 1b, we compare 1-particle and 2-particle MR ($2a = 1.16$ μm probe particle, probe particle distance $R = 8$ μm) carried out in a densely cross-linked actin network (w/o myosin). In contrast to what was observed in entangled actin networks, $G(\omega)$ estimated from 1-particle and 2-particle MR here agreed. This result indicates that the cross-linked actin network can be approximated as a homogeneous continuum on length scales larger than the micron-scale probes used here.

Having established that the mechanical response of the cross-linked actin cytoskeleton (w/o myosin) is not length-scale dependent, we proceed to analyze how myosin generates forces, and how the actin network responds to those. In general, an active force generator can be expanded into a sum of force multi-poles, $f \varepsilon^n \nabla^n \delta(r)$, where $n$ is the multipole order (monopole: $n = 0$, dipole: $n = 1$, etc.), and $\varepsilon$ and $f$ are microscopic length and force scales of the force generator, respectively. Flows in viscous media and displacements in elastic media created around a force multipole are long-ranged and decay as power-laws, scaling as $1/r^{n+1}$. When active forces are generated randomly and independently by many force generators distributed in a homogeneous and isotropic viscoelastic medium, non-thermal cross-correlations between the fluctuations of two embedded probe particles are independent of particle size, but scale as $1/R^{2n-1}$, with $R$ being the particle distance [34]. In contrast, the square of the autocorrelation of the fluctuations (power spectral density) of an individual probe scales as $1/a^{2n-1}$, with $a$ being the probe radius. As shown in Fig. 2, our results support the force dipole model ($n = 1$, see supplementary materials) as assumed in prior works [4, 35]. This is the lowest-order multipole consistent with internal force balance.

Non-thermal fluctuations in reconstituted actin gels activated by muscle myosin mini-filaments typically appear when the concentration of ATP has decreased sufficiently after sample preparation [4]. Muscle myosin minifilaments can crosslink and slide two actin filaments in the network only at relatively low ATP concentrations, when the duty ratio (on- vs. off-time) of myosin heads is large enough [4, 36, 37]. The amplitude of non-thermal fluctuations in the gels reach a maximum just before super-precipitation [28] occurs, at which point the stresses generated by myosin motors are sufficient to collapse the network and cause phase separation. Before super-precipitation, probe particle pairs show distinct fluctuation correlations that depend on the specific geometrical arrangements of the particle pairs and the (small) number of myosin molecules in their vicinity [4].

We therefore averaged correlations for N = 378 particle pairs ($2a = 1.1$ μm, polystyrene beads) following the protocol described below. As discussed later, in Fig. 3a, the power spectral density of non-thermal 2-particle correlations in actin-myosin gels exhibits a power-law dependence in the form of $C^{12}(\omega) \sim \omega^{-2}$. The strength of correlations also depends on the probe separation $R$. Therefore, we took the average of $B^{12} \equiv \omega^2 C_{||}^{12}(\omega)$ for pairs with similar separations $R$ (Fig. 2a). We clearly observe $1/R$ dependence of $\omega^2 C_{||}^{12}(\omega)$ as shown by the solid line, whereas a $1/R^3$ power law was not consistent with the data in the range of $R$ observed. Note that for thermal fluctuations, in equilibrium, the same distance dependence ($\propto 1/R$) of 2-particle correlations has been reported [12]. In the case of non-equilibrium fluctuations, however, the $1/R$ dependence is not generally expected as we discussed above, but it was observed here.

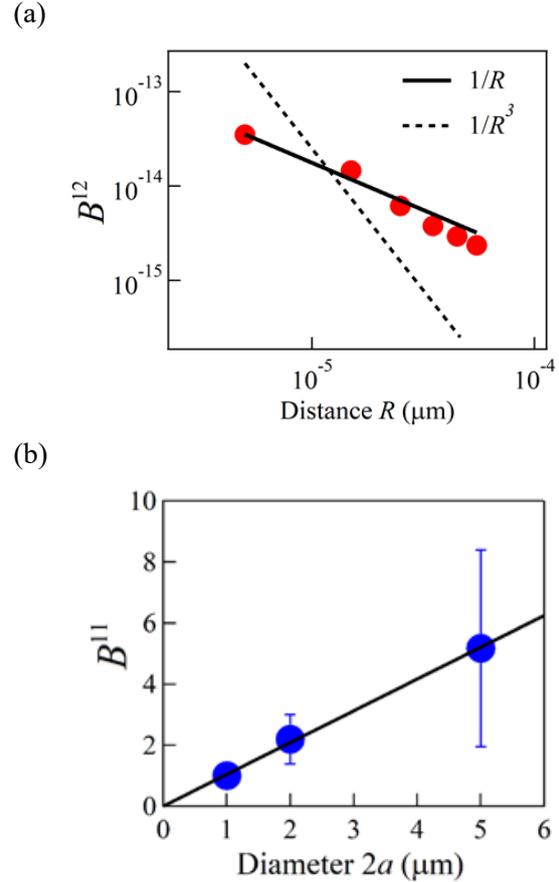

Fig. 2: (a) Non-thermal 2-particle parallel fluctuation correlations normalized as $B^{12} \equiv \omega^2 C_{||}^{12}(\omega)$. The solid and broken lines show $1/R$ and $1/R^3$ dependence, respectively. (b) Particle size dependence of 1-particle fluctuations. 1, 2, and 5 μm probe particles are dispersed in the same active gel samples. Non-thermal fluctuations at low frequencies ($0.1 < \omega/2\pi < 1$) are fitted with the power-law function $C^{11}(\omega; 2a) = A_{2a} \omega^x$, with the exponent $x$ fixed to the value obtained for $2a = 1$μm. Normalized correlations $B^{11} \equiv C^{11}(\omega, 1\mu m)/C^{11}(\omega, 2a)$ are then obtained. The results are averaged for 6 different samples.

Fig. 2b shows the particle size dependence of 1-particle autocorrelation amplitudes. The strength of non-thermal fluctuations depended on the elapsed time after sample preparation (and thus the remaining concentration of ATP) [4, 36]. To be able to test different probe sizes simultaneously, probe particles with several different sizes (diameter $2a$ = 1, 2, and 5 μm) were therefore dispersed in the same sample, and the ratio of power spectra $C^{11}(\omega; 2a)$ to that of probe particles with $2a = 1\,\mu\text{m}: C^{11}(\omega; 1\mu\text{m})$, i.e. $B^{11} \equiv C^{11}(\omega; 1\mu\text{m})/C^{11}(\omega; 2a)$, was calculated for particles observed in the same field of view (Fig. 2b). We find that $B^{11}$ linearly depends on particle size $a$ [22], which again supports the force dipole model.

Fig. 3a shows results of 2-particle AMR and PMR in cross-linked active gels measured using both video microscopy and laser interferometry. Laser interferometry was carried out for a single particle pair (distance $R$ = 6 μm), and results were time-averaged for approximately 1 h. Having confirmed that both thermal and non-thermal cross-correlations have a particle distance dependence of $1/R$, averages for different particle pairs were calculated and normalized to the value for $R$ = 6 μm, as $\langle C^{12}_{\|(\perp)}(\omega, R=6\mu\text{m})\rangle = \{\sum_{i,j} R^{ij} C^{ij}_{\|(\perp)}(\omega)\}/(\text{N} \cdot 6\times 10^{-6}[\text{m}])$. 2-particle AMR could not be conducted in the active gels because the large non-equilibrium fluctuations resulted in a small signal-to-noise ratio. Therefore, 2-particle AMR results were simulated by normalizing the value of 1-particle AMR as $A^{12}_{\|}(\omega) = 3A^{11}(\omega)a/2R$ based on Eq. (2). Here we used the fact that the response of cross-linked actin gels was found to be length-scale independent (Fig. 1b). At frequencies lower than 100 Hz, we again observe the violation of FDT in 2-particle correlations, while AMR and PMR are consistent at higher frequencies, indicating that thermal fluctuations dominate. Small differences at high frequencies are most likely caused by fluid inertia that becomes measurable for high-frequency correlations in 2-particle MR [21, 38]. Note that 2-particle AMR was estimated from 1-particle AMR data for which the inertial dampening of the response is smaller.

In Fig. 3b, we plot the ratio of the fluctuation cross-correlations measured for the two independent modes in 2-particle PMR, $C^{12}_{\perp}/C^{12}_{\|}$, as a function of frequency. At high frequencies where thermal fluctuations are dominant, we find $C^{12}_{\perp}/C^{12}_{\|} = 1/2$ as expected for an incompressible medium from Eqs. (1) and (2). The downturn of $C^{12}_{\perp}/C^{12}_{\|}$ at frequencies higher than 10 kHz is quantitatively understood due to the frequency dependence of fluid inertial dampening that appears differently in the parallel and perpendicular modes [21, 39]. In contrast, at frequencies lower than 10 Hz, where non-thermal fluctuations are dominant, a conspicuously different ratio of $C^{12}_{\perp}/C^{12}_{\|} \sim 1$ was observed. This anomalous ratio is not immediately understandable and is not expected, even for the predominantly non-thermal fluctuations. In an isotropic homogeneous incompressible continuum, non-thermal fluctuations generated by randomly-distributed force dipoles would give rise to 2-particle correlations [34]

$$C^{12}_{\|}(R,\omega) = \frac{\langle c\kappa^2_{dip}\rangle}{60\pi R |G|^2}, \quad C^{12}_{\perp}(R,\omega) = \frac{\langle c\kappa^2_{dip}\rangle}{120\pi R |G|^2}. \quad (3)$$

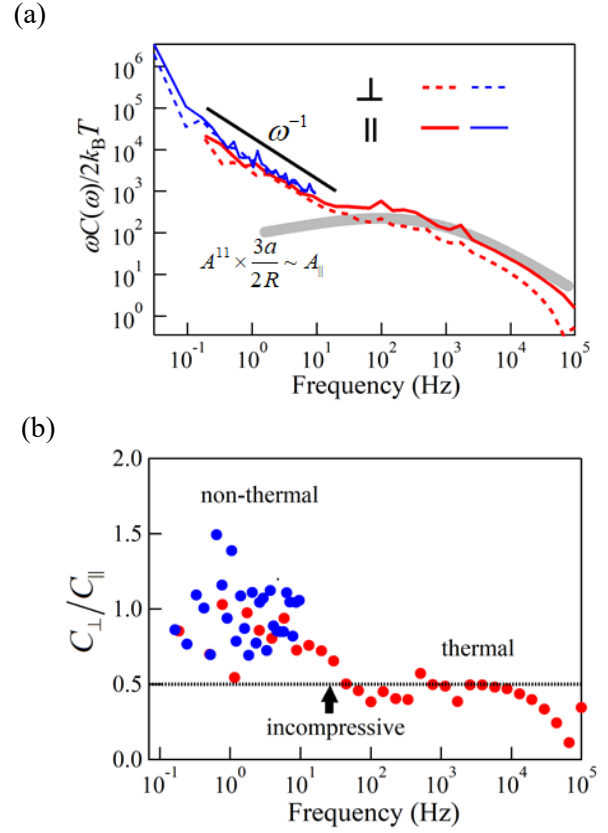

Fig. 3: (a) 2-particle parallel and perpendicular correlation spectra $\omega C(\omega)/2k_B T$ observed by both video microscopy (blue lines, normalized to $R$ = 6 μm) and laser interferometry (red lines). Continuous and dotted lines show parallel and perpendicular directions, respectively. The grey band indicates the parallel thermal correlations estimated from 1p AMR as $A^{11} \times (3a/2R) \sim A_{\|}$. (b) Ratio of the cross-correlation spectra in parallel and perpendicular directions $C_{\perp}/C_{\|}$. Thermal fluctuations at frequencies higher than 100 Hz give the theoretically predicted value for an incompressible network $C_{\perp}/C_{\|} \sim 1/2$, while non-thermal fluctuations at lower frequencies give an anomalous value $C_{\perp}/C_{\|} \sim 1$, which is due to the local nonlinear response of actin filaments.

Here, $c$ is the number density of myosin mini-filaments (~ $10^{16}$ /m$^3$), $\kappa_{dip} \equiv \varepsilon f$ is the moment of a single force dipole, and linear response of the medium is assumed. Eq. (3) shows that the ratio $C^{12}_{\perp}/C^{12}_{\|}$ is still expected to be 1:2. The assumptions made to derive Eq. (3) are mechanical isotropy, homogeneity, and length-scale-independent response. These assumptions have been confirmed for cross-linked actin networks at length scales larger than μm.

Although hydrogels are incompressible when deformed as a whole by external forces, except at very long time scales, networks can, in principle, be locally compressed by draining solvent at the expense of swelling neighboring regions. Such solvent-network decoupling has been quantitatively discussed for linearly elastic networks [9, 10, 34, 40, 41], and found possible only for very slow fluctuations at microscopic length scales. Decoupled (compressive) fluctuations in a two-component linearly elastic continuum are possible at frequencies less than

$$f_c \sim \xi^2 G / 2\pi\eta R^2, \qquad (4)$$

where $\xi$ is the network mesh size, $\eta$ is the solvent viscosity, and $R$ is the characteristic length of the fluctuation [10, 41]. For our 2-particle MR experiments, $f_c$ is estimated to be $\sim 10^{-2}$ Hz, taking $R$ to be the probe separation. This estimate does not support the idea that our 2-particle MR data obtained at higher frequencies are influenced by compressive fluctuations. Furthermore, when network and solvent are decoupled, a Poisson ratio of $\nu_{dec} = 1/4$ is expected for linear and affine network deformations [34]. Therefore, even if experiments were performed at frequencies lower than $f_c$, the maximum theoretical value of $C_\perp^{12}/C_\parallel^{12}$ is estimated to be 0.66 [34], which is still smaller than the experimentally observed value. Thus, as long as linear response is assumed, $C_\perp^{12}/C_\parallel^{12} \sim 1/2$ is expected in the whole measured frequency range for our active gel sample.

To explain our contrasting findings, we will question the validity of the assumption that the network (locally) responds linearly to motor-generated forces. One of the hallmarks of semi-flexible polymers such as actin filaments, is their highly nonlinear mechanical response [42-44]. For polymers whose contour length $L$ is similar or shorter than their persistence length $l_p$, the end-to-end distance $x$ under an applied tension $\sigma$ can be expressed as [45, 46]

$$x(\sigma) = L - \frac{k_B T}{2\sigma}\left(L\sqrt{\frac{\sigma}{l_p k_B T}} \coth\left(L\sqrt{\frac{\sigma}{l_p k_B T}}\right) - 1\right). \qquad (5)$$

In Fig. 4a, we plot the force extension curve of an actin filament ($L = 2$ μm, $l_p = 10$ μm) which highlights the very abrupt stiffening response under stretching, while the filament is weak to compression. Myosin mini-filaments typically generate forces on the scale of tens of pN. The response of actin filaments in a myosin-activated network is thus expected to be highly nonlinear and asymmetric around the equilibrium end-to-end distance ($\sim 1.93$ μm) at $\sigma = 0$. Actin filaments that directly interact with myosin during contraction should be compressed and buckled in the direction normal to the force dipole axis as schematically shown in Fig. 4b. Note that such microscopic compression does not appear for linearly-responding filaments because of the symmetry between stretching and compression [25].

Around the active myosin motor, the local contractile stress would predominantly propagate along the axis of the dipole to distances longer than individual actin filaments because of the stiffening of the network along that direction [24, 47]. On the other hand, compressive stresses and strains normal to and in the plane of the dipole do not propagate far in the normal direction because of the buckling of the corresponding filaments. Along the axis of the dipole, however, off-axis compression is expected to extend to distances beyond the actin filaments to which the active myosin is directly bound (Fig. 4b) because of the stiffening of filaments under tension. The collective multi-filament response thus consists of axial contraction and lateral compression, i.e. a local concentration increase or network compression. At larger distances, propagation of the contractile stress is expected to diffuse in direction and become more isotropic [25].

So far we have not taken into account viscous drag between filaments and solvent. The described mechanism of local compression implies network motion to be decoupled from the surrounding solvent and is therefore expected to occur at long times or low frequencies. To estimate at what frequency compression could occur, we can, however, not use Eq. (4) because of the "nonlinear" response of the filaments. The strain hardening of the stretched filaments acts like a strongly increased modulus $G$ in Eq. 4 leading to slipping of the filament against the surrounding solvent Therefore the decoupling frequency is pushed higher than the estimate in Eq. (4). The contribution of local nonlinear compressions extending towards mesoscopic length scales was not considered when deriving Eq. (3). Note that a similar isotropic compression, referred to as contractile foci, has been observed during slow, irreversible aggregation of acto-myosin networks where presumably many myosins act in concert to overpower crosslinks and contract the network [48].

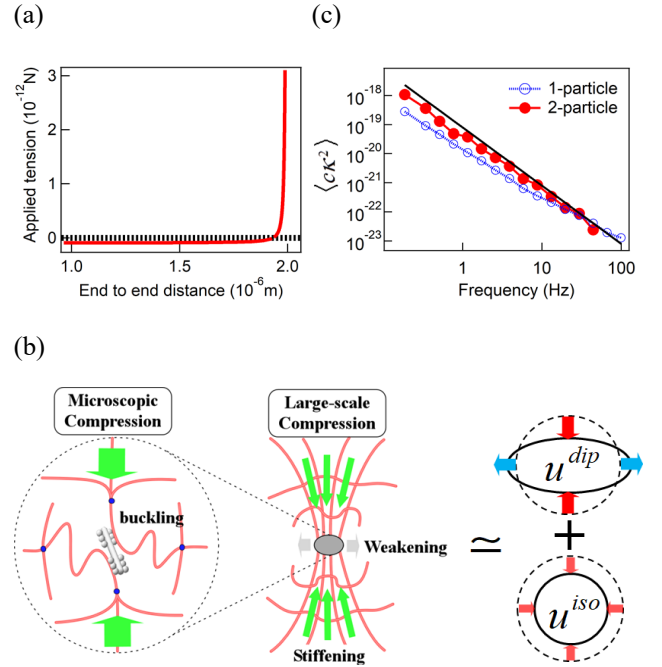

Fig. 4: (a) Theoretical force-distance curve of an actin filament. (b) Schematic representation of the microscopic and larger scale compressions induced by nonlinear response of actin filaments to myosin's dipolar forces. Large-scale compression was approximated as a sum of $u^{dip}$ and $u^{iso}$. (c) Power spectral density of force dipole fluctuations $\langle c\kappa^2 \rangle$ calculated in Eqs. (8) and (9) in the text for 1-particle (blue open circles) and 2-particle MR (red filled circles), respectively. The solid line shows a $\omega^{-2}$ power law.

In lieu of a full nonlinear continuum description of the network deformation, we incorporate the nonlinear effect as an isotropic compression added to the ordinary linear response. The deformation field is approximated by superimposing two fields: (i) $u^{dip}$ due to a single force dipole $\kappa_{dip}$, and (ii) $u^{iso}$ due to the isotropic compression $\kappa_{iso}$. In ref. [34], we have calculated the correlations of 2-particle fluctuations ($C_\perp^{12}$ and $C_\parallel^{12}$) that will appear when $\kappa_{dip}$ and $\kappa_{iso}$ are randomly distributed in elastic homogeneous media.

Since $u^{dip}$ is the linear-response contribution to the probe displacements, the coupling of the network to the solvent in this case will ensure incompressible response ($\nu = 1/2$) of the non-local displacements. Eq. (3) then applies to the probe correlations driven by $u^{dip}$. In contrast, $u^{iso}$ is approximated as described in Eq. S4, using an apparent Poisson ratio $\nu^{NL} < 1/2$ that describes the compression due to the nonlinear response. Note that this approximation is consistent with the observed probe cross-correlations that scale as $1/R$. It has been shown that $u^{iso}$ does not produce any correlation in parallel fluctuations ($C_\parallel^{12} = 0$), but only contributes to the perpendicular component as [34]

$$C_\perp^{12}(R,\omega) = \frac{\langle c\tilde{\kappa}_{iso}^2 \rangle}{32\pi R |G|^2}, \qquad (6)$$

where $\tilde{\kappa}_{iso} \equiv \{(1-2\nu^{NL})/(1-\nu^{NL})\}\kappa_{iso}$. Considering both contributions, $u^{dip}$ and $u^{iso}$ [Eqs. (3) and (6)], we obtain for 2-particle correlations:

$$C_\parallel^{12}(R,\omega) = \frac{\langle c\kappa_{dip}^2 \rangle}{60\pi R |G|^2},$$

$$C_\perp^{12}(R,\omega) = \frac{\langle c\kappa_{dip}^2 \rangle}{120\pi R |G|^2} + \frac{\langle c\tilde{\kappa}_{iso}^2 \rangle}{32\pi R |G|^2}. \qquad (7)$$

Here the cross-terms between $u^{dip}$ and $u^{iso}$ vanish under orientational averaging because of the orientational symmetry of isotropic compressions. The emergence of $u^{iso}$ due to nonlinear isotropic compression can thus explain our experimental observation $C_\perp^{12}/C_\parallel^{12} \sim 1$ if $\tilde{\kappa}_{iso} \sim \kappa_{dip}/2$.

The total myosin activity should be written as a combination of $\kappa_{dip}$ and $\tilde{\kappa}_{iso}$, although the phenomenological nature of $\tilde{\kappa}_{iso}$ obscures the exact formulation. For simplicity, we here assume $\kappa \equiv \kappa_{dip} + \tilde{\kappa}_{iso}$. Note that a different weighting of the components merely leads to proportional shifts of the results [Eq. (8) and (9) below]. To check for consistency, we can now estimate the dipole strength in two different ways, from 2-particle and from 1-particle fluctuations. The observed 2-particle correlations can, with the above assumption, be expressed as

$$C_\parallel^{12}(R,\omega) \sim C_\perp^{12}(R,\omega) \sim \langle c\kappa^2 \rangle / 135\pi R |G|^2. \qquad (8)$$

The 1-particle non-thermal power spectral density $C^{11}(\omega)$ can be expressed as (see supplementary information)

$$C^{11}(\omega) \sim \frac{\langle c\kappa^2 \rangle}{138\pi a |G^2|}. \qquad (9)$$

Fig. 4c shows that $\langle c\kappa^2 \rangle$ estimated both ways using Eqs. (8) and (9) are approximately consistent.

## 4. Discussion

Dynamic tensions generated by the contractile acto-myosin cortex are fundamental for cell physiology, development, and homeostasis [2]. A contractile system with the distinctly asymmetric response of semiflexible filaments such as actin, is, however, in danger of collapsing and phase separating. In reconstituted acto-myosin model systems this indeed occurs, and a dramatic macroscopic collapse of the active network and irreversible phase separation is typically observed, referred to as superprecipitation [28]. In living cells, in contrast, tensions and contractile force generation as well as crosslinking, filament polymerization and depolymerization dynamics are carefully regulated to prevent run-away catastrophes, and superprecipitation is not usually observed. There is, however, evidence that active concentration fluctuations in the acto-myosin cortex are responsible for patterning membrane proteins and are crucial for the function of receptors and signaling molecules [49-51]. Our *in vitro* study shows that local, rapid acto-myosin contractions can occur reversibly in active gels due to the strong non-linearity of filament response. Large-scale phase separation is, in our case, frustrated by the presence of crosslinks. The mechanical integrity and large-scale homogeneity of the network are therefore maintained. Such a dynamic steady state of an active contractile network appears to be closely related to actual cell physiology.

Up to now, we have focused on local compressive network dynamics explaining the observed anomaly in the ratio of fluctuation correlations. In the following, we discuss several issues that are related to this main result of our study. First, we comment on the length-scale dependences observed in the single-particle fluctuation auto-correlations and 2-particle fluctuation cross-correlations in active gels ($C^{11} \propto 1/a$ and $C^{12} \propto 1/R$). Here, we attribute these length-scale dependences to the continuum mechanics of the system that transmits stresses and strains created by the active force dipoles over long distances. The displacement (velocity) field around an individual dipole in an elastic (viscous) medium decays as $1/r^2$ with distance $r$ from the force dipole. The sum of such fields created by multiple force dipoles randomly distributed in a sample then explains $C^{11} \propto 1/a$ and $C^{12} \propto 1/R$.

Confusingly, thermal fluctuations of probe particles in isotropic viscoelastic continua show the same scaling behaviors [ $C^{11} \propto 1/a$ , $C^{12} \propto 1/R$ ; see Eqs. (2) and (1)]. In this case, however, the origin of thermal forces (dynamic interactions between microscopic objects such as collisions between solvent molecules or solvent molecules and probe particle) do not directly act over large distances. Although it is not possible to measure the microscopic dynamics that are fast and nonlinear, their collective effect appears as the fluctuating monopole force driving probe motions that results in $C^{11} \propto 1/a$. The probe motion then creates long-ranged fluctuations that decay as $1/r$ with the separation from the force monopole. Thus, the power-law dependency $C^{12} \propto 1/R$ for thermal correlations is caused by the "monopole" field created around probe particles. Non-thermal 1-particle fluctuations and 2-particle correlations were previously analyzed by extending this classical thermal model and derived the same $1/a$ and $1/R$ dependence, respectively [1], but with a totally different basis from the dipole model.

Which of these two models applies to a specific non-equilibrium systems depends on the time scale of the activity and how far the fluctuations propagate in the medium. When active force generation merely creates molecular-scale correlations in both space and time, the prior (monopole) model is appropriate. In our actin gels activated by myosin motors, however, motors directly drive long-ranged and long-term correlations [4,36] Therefore, the dipole model of active myosin is appropriate for non-thermal fluctuations in these active gels.

It is appropriate to model a probe particle in an equilibrium system as a force monopole when considering correlated thermal fluctuations sufficiently far from the particle. Correlated thermal fluctuations close to the particle, however, can exhibit higher-order contributions. Recently, Segev et al. [52] interpreted their observation of enhanced correlations between probe pairs in thermal hydrogels at relatively small separations as higher-order contributions ($n = 2$ with $C^{12} \propto 1/R^3$). It is worth to discuss the conditions under which these unusual correlations (that were not observed in our results) are observed. In Segev's experiments, the probe particles were coated with a polymer brush to avoid direct adherence to the network [52]. Probe fluctuations are in that case expected to be enhanced because they are linked to the surrounding solvent but somewhat decoupled from the network due to the extended buffer layer around the particles. The solvent fluctuations that accompany the enhanced probe motions penetrate into the surrounding gel over the length of typically several μm [52], but eventually couple to network fluctuations and decay normally beyond that length. In our experiments, however, the probe surfaces were not coated with a polymer brush. Most importantly, the dominant forces are generated by the motor proteins that directly interact with the actin filaments making up the network. The forces then propagate via the network to the probes. Furthermore, any higher-order multipole contribution should rapidly decay around molecular motors that have microscopic dimensions. For all of these reasons, we did not expected to observe higher-order contributions to 2-particle correlations in our experiments.

We further need to discuss the non-equilibrium relation between motor-driven non-thermal fluctuations and the increase of active gel elasticity they cause. We can estimate the active tension applied to individual actin filaments in active gels from the observed fluctuations. As has been reported [4,36,37], the forces generated by myosin mini-filaments are typically creating a slow stress increase followed by rapid release due to the avalanche-like detachment from the actin network. Because of this step-like release, we expect a frequency dependence of $\langle c\kappa^2 \rangle \sim \omega^{-2}$ as is actually observed over a wide range of frequencies [22,40,53]. At frequencies lower than the inverse of the average time, over which a myosin mini-filament stays bound and exerts force, we expect a cross-over to $\langle c\kappa^2 \rangle \sim \omega^0$ [34].

In Fig. 5a, we show the stress fluctuation spectrum measured with a 5 μm probe particle (1-particle MR) showing a clear crossover at around $f_c \sim 0.5$ Hz, but not to a plateau, but rather to a weaker power law, possibly due to the structural relaxation (creep response) of the gel. We did not see any such crossover when smaller probes ($2a \sim 1$ μm) were employed (Fig. 3a). The fact that there was no crossover for small particles is likely due to hysteresis observed in probe movements in active gels; after strong force generation, the particle may not return to its original position. Such hysteresis should become larger when probe particles are similar or smaller than the cross-linking distance $l_c \sim $ μm. We therefore estimate the average strength $\bar{\kappa}$ of force dipoles by integrating the power spectral density of force fluctuations above the crossover frequency
$$c\bar{\kappa}^2 = f_c \langle c\kappa^2 \rangle \Big|_{f=f_c} + \int_{f_c}^{\infty} \langle c\kappa^2 \rangle df \sim 10^{-18} \text{ [J}^2/\text{m}^3\text{]}.$$
Under the continuum mechanics approximation, strains around a force dipole $\bar{\kappa}$ are readily calculated. Assuming affine deformation of the network, we can then obtain an estimate of filament extensions as a function of distance from the force dipole and filament orientation. In a prior study, we have estimated the averaged tension $\sigma_0$ by integrating over a random distribution of actin filaments as: [34]

$$\sigma_0 \sim \frac{l_p^2 k_B T c \bar{\kappa}}{l_c^3 G}, \quad (10)$$

which results in a tension per filament of a $\sim 10^{-2}$ pN from the data shown in Fig. 4a (with $\bar{\kappa} \sim 10^{-18}$ J, $c \sim 10^{18}$ m$^{-3}$, $l_p^2 \sim 10^{-10}$ m$^2$, $l_c^3 \sim 10^{-17}$ m$^3$, $G \sim 10$ Pa).

An independent estimate of $\sigma_0$ in the active gel can be obtained from the active stiffening of the network that affects both plateau and high frequency modulus. The internal tension applied to the actin network by active myosins increases the plateau value of the network elasticity $G^0$ [4]. The network elasticity originates from the resistance of individual filaments to elongation/compression imposed by external macroscopic deformation. Under the assumption of affine deformation, the macroscopic elasticity of a cross-linked network is thus obtained by averaging the response of individual filaments. The stiffness of a filament subjected to tension $\sigma_0$ is obtained as $d\sigma/dx|_{\sigma=\sigma_0}$ using the force-extension relation $x(\sigma)$ in Eq. (5). From this, the (frequency independent) plateau shear modulus $G^0$ can be estimated as: [54, 55]

$$G^0 \sim \frac{1}{15} \rho l_c \left.\frac{d\sigma}{dx}\right|_{\sigma=\sigma_0} \quad (11)$$

where $\rho$ is the length density of actin filaments per volume with units of inverse length squared. However, the elastic plateau of our active gels not completely flat, but in practice, has a residual frequency dependence. The tension $\sigma_0$ is better estimated from the high-frequency behavior of the response function $A''$ according to: [4]

$$A''(\omega) = \text{Im}\left[\frac{5}{2\pi a \rho l_c} \sum_q \frac{2 k_B T q^4}{\omega_q \zeta^2 (2\omega_q - i\omega)}\right], \quad (12)$$

where $\omega_q = (\sigma_0 q^2 + \kappa q^4)/\zeta$ accounts for both stretching and bending energy of filaments, $q = n\pi/l_c$, $n = 1, 2, 3, ....$, and $\zeta$ is the transverse friction coefficient per unit length of a filament. By fitting Eq. (12) to the high frequency parts of the observed response functions, we obtain an estimate for $\sigma_0$.

In Fig. 5b, we show the increase of $G^0$ ($G'$ at 1.6 Hz) with the applied tension in different preparations of active acto-myosin gels. The curve is the theoretical one based on Eq. (11). The data point marked by an arrow ($\sigma_0 \sim 0.1$ pN) is the one taken from the sample shown in Fig. 5a. The analysis of non-thermal probe fluctuations using Eq. (10) seems to somewhat underestimate the quantity of $\sigma_0$ ($\sim 10^{-2}$ pN). The reason for this phenomenon might be collective effects. As we have discussed above, active stresses rapidly propagate to large distances because of the nonlinear response of actin filaments. The stresses from multiple different motors could thus add up to provide the average tensile stress on filaments. Although this steady tension would not contribute to the fluctuations of probes, it could contribute to overall compression and stiffening in the gel [56, 57]. Nevertheless, it is remarkable that the tension obtained from the change of the mechanical response (Fig. 5b) is on the same order of magnitude as the independent estimate from the scaling argument using the non-equilibrium stress fluctuations (Fig. 5a).

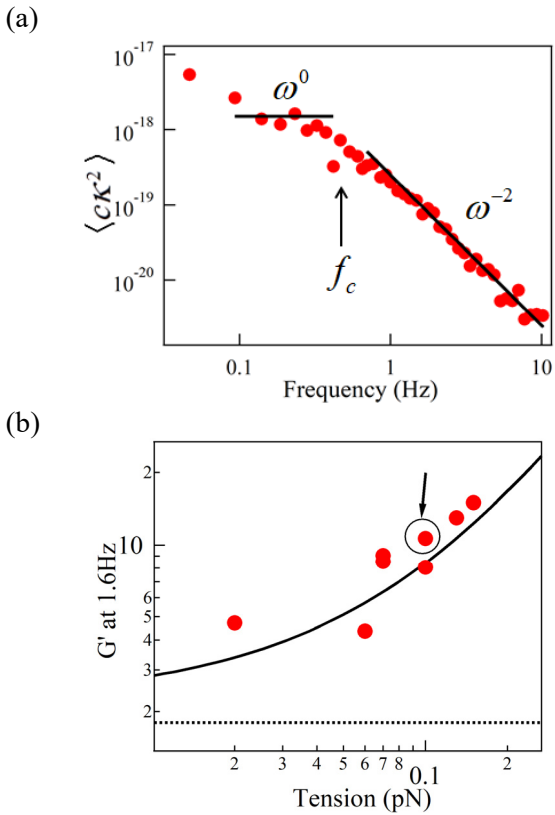

Fig. 5: (a) Power spectral density of force dipole fluctuations $\langle c\kappa^2 \rangle$ measured with 1-particle MR using probe particles ($2a = 5$ $\mu$m) larger than typical cross-linking distances. The crossover from high frequency $\omega^{-2}$ to low frequency behavior was observed around $f_c$ that indicates the time scale for the processive force generation by myosin mini-filaments. The solid lines are to guide the eye, showing $\omega^{-2}$ and $\omega^0$ power laws. (b) Relation of the elastic plateau $G^0$ (estimated as $G'$ at 1.6Hz) with tension applied to single actin filaments (estimated by fitting Eq. (12) to the high frequency part of $A''$). The symbol marked by an arrow is obtained from the same set of data used in (a). The drawn line indicates the theoretical prediction of Eq. 11.

## 5. Conclusions

We have investigated non-thermal fluctuations in actin/myosin active gels using active/passive and 1-particle/2-particle microrheology techniques. We quantified the 1-particle fluctuations and 2-particle correlations of motor-generated non-thermal fluctuations in cross-linked actin networks. The length-scale dependence (probe-size dependence for 1-particle fluctuations and probe-separation dependence for 2-particle correlations) of non-thermal fluctuations conformed to the theoretical predictions based on the distributed dipole model of force generators [34].

Surprisingly, fluctuations of pairs of particles were much more strongly correlated than expected normal to the direction connecting the particles. Going beyond linear response theory in the force-dipole model, we attribute the observed anomaly to a strongly nonlinear response of actin filaments (buckling for compressive and stiffening in tensile direction) that leads to rapid but reversible local network draining. The microscopic compression of the network rapidly extends at least to several micrometers because active stress can efficiently propagate along the stiffened actin filaments while the viscous coupling to the surrounding solvent becomes less dominant. The tension (pre-stress) imposed on the network by the embedded force-generating myosins then also causes global stiffening of the network. The highly-nonlinear response of semi-flexible polymer networks to molecular force generators thus consistently explains both experimentally observed phenomena: anomalous non-equilibrium fluctuations and active stiffening of the network.

## Conflicts of interest

There are no conflicts to declare.

## Acknowledgements

We acknowledge discussions with D. A. Head, and experimental supports by T. Toyota. This work was supported by JSPS KAKENHI Grant Number JP18H01189, JP25103011, JP15H03710 (to DM). This research was supported in part by the European Research Council under the European Union's Seventh Framework Programme (FP7/2007- 2013) / ERC grant agreement n°340528 (to CFS)